\documentclass[conference,compsoc]{IEEEtran}
\ifCLASSOPTIONcompsoc
  \usepackage[nocompress]{cite}
\else
  \usepackage{cite}
\fi
\ifCLASSINFOpdf
   \usepackage[pdftex]{graphicx}
\else
   \usepackage[dvips]{graphicx}
\fi
\usepackage{algorithmic}
\ifCLASSOPTIONcompsoc
  \usepackage[caption=false,font=footnotesize,labelfont=sf,textfont=sf]{subfig}
\else
  \usepackage[caption=false,font=footnotesize]{subfig}
\fi
\usepackage{url}
\usepackage{hyperref}
\hypersetup{
    colorlinks=true,
    linkcolor=blue,
    filecolor=magenta,      
    urlcolor=cyan,
    pdfpagemode=FullScreen,
    }
\usepackage{multirow}

\begin{document}
\textbf{This preprint version of the paper was submitted to IEEE BIBM24 Conference. The final version can be found in the proceedings of the conference.}

\title{Combining knowledge graphs and LLMs for hazardous chemical information management and reuse}

\author{\IEEEauthorblockN{Marcos Da Silveira, Louis Deladiennee, Kheira Acem, and Oona Freudenthal}
\IEEEauthorblockA{Luxembourg Institute of Science and Technology\\5, avenue des Hauts-Fourneaux L-4362 Esch-sur-Alzette, Luxembourg\\
Email: \{marcos.dasilveira, louis.deladiennee, oona.freudenthal\}@list.lu}
}

%


\maketitle

\begin{abstract}
Human health is increasingly threatened by exposure to hazardous substances, particularly persistent and toxic chemicals. The link between these substances, often encountered in complex mixtures, and various diseases are demonstrated in scientific studies. However, this information is scattered across several sources and hardly accessible by humans and machines. This paper evaluates current practices for publishing/accessing information on hazardous chemicals and proposes a novel platform designed to facilitate retrieval of critical chemical data in urgent situations. The platform aggregates information from multiple sources and organizes it into a structured knowledge graph. Users can access this information through a visual interface such as Neo4J Bloom and dashboards, or via natural language queries using a Chatbot. Our findings demonstrate a significant reduction in the time and effort required to access vital chemical information when datasets follow FAIR principles. Furthermore, we discuss the lessons learned from the development and implementation of this platform and provide recommendations for data owners and publishers to enhance data reuse and interoperability. This work aims to improve the accessibility and usability of chemical information by healthcare professionals, thereby supporting better health outcomes and informed decision-making in the face of patients exposed to chemical intoxication risks.  
\end{abstract}


%
\IEEEpeerreviewmaketitle

\section{Introduction}
In today's industrialized world, exposure to hazardous chemicals is an ever-present risk, both in occupational settings and through environmental contamination. These substances, widely used across various industries, pose significant health risks to individuals, necessitating rapid and accurate access to detailed chemical information by healthcare professionals. For many years, governments and organizations have collaborated to create efficient systems for sharing digital information about hazardous chemicals. They have enacted laws, established rules, and developed programs to define and implement governance strategies, standards, services, and infrastructures aimed at creating a more effective data market. Notable examples in Europe include the Data Governance Act \cite{DGA}, the Data Act \cite{EUDA}, the Zero Pollution Action Plan \cite{zeroPAP}, Chemicals Strategy for Sustainability \cite{CSS2020}, and the Waste Framework Directive  \cite{EUWFD} designed to facilitate information sharing. Additionally, international initiatives such as the \href{https://rd-alliance.org/}{Research Data Alliance} (RDA), formed in 2013 by the EU, the USA, and Australia, advise organizations on requirements for interoperable and reusable data.

However, relevant information about chemical substances remains difficult to access. It is scattered across several websites, stored in various formats, and often lacks metadata and indexes. To access the information, users must visit diverse websites, enter queries in their browsers, or open and read CSV files. The data format and the meaning of fields in each file are defined by the data publisher and often not documented in a computer-interpretable format. Although initiatives like \href{https://www.epa.gov/assessing-and-managing-chemicals-under-tsca/introduction-chemview}{ChemView} or \href{http://www.chemspider.com/}{ChemSpider} add value by aggregating information from various sources into a single website, the interpretation, comparison, and linking of diverse information still require human intervention. This makes it challenging, for instance, to identify the relationship between substances and the diseases diagnosed in patients. 

The first contribution of this paper is the analysis of widely adopted data sources on hazardous chemical substances and evaluate their compliance with the principles of Findability, Accessibility, Interoperability, and Reusability (FAIR)\cite{wilkinson2016fair}. 

The second contribution is the analysis of how the knowledge available in several data sources can be linked and exploited. We propose to investigate the use of Knowledge graphs (KGs) to represent the connection between hazardous substances and diseases. KGs are particularly well adapted to represent links between diverse pieces of information \cite{da2024knowledge}. They excel at integrating data from multiple sources, ensuring that relationships between various entities, such as chemical substances and associated diseases, are clearly defined and easily navigable. This structured representation makes it easier to discover connections and derive insights that might be missed when data is dispersed across different formats and locations. For instance, in the context of hazardous chemical substances, a KGs can link chemical properties, products/usage contexts, organs often affected by the substance, and documents with relevant information. This integrated view supports healthcare professionals in quickly identifying relevant data, understanding the potential health risks associated with exposure, and making informed decisions about patient care. Additionally, KGs enhance data interoperability and reuse, as they provide a standardized format for representing and querying complex datasets. By facilitating automated reasoning and advanced analytics, KGs contribute to more efficient and accurate data analysis, ultimately improving the accessibility and utility of critical information in fields such as toxicology and public health.

However, the adoption of KGs technology in the healthcare sector remains marginal. Querying these graphs requires specialized ICT knowledge, which is often lacking among healthcare professionals. The recent advances in large language models (LLMs) and its chatbot version are creating opportunities to overcome this barrier, allowing for more intuitive and user-friendly interactions with complex datasets. Chatbots powered by LLMs can act as intermediaries, interpreting natural language queries from healthcare professionals and translating them into precise queries for KGs. This development makes it feasible for doctors, nurses, and other medical staff to access and utilize comprehensive data without needing advanced technical skills.

For instance, a doctor could ask a chatbot in a natural language \textit{"What are the potential health impacts, particularly on the heart, of exposure to Acrylaldehyde?"}, and the chatbot could retrieve and present relevant information from a KG where links between chemical substances (e.g. Acrylaldehyde) and diseases related to the heart (e.g., heart valve disease, heart-hand syndrome, slovenian type, etc.) are represented. It can also provide links to documents where deeper discussion can be found. This seamless interaction not only saves time but also enhances the accuracy of information retrieval, thereby improving patient care. By bridging the gap between complex data structures and everyday medical practice, LLMs and chatbots are poised to significantly enhance the adoption and utility of KGs technology in healthcare.

This paper is structured as follows: first, we provide an overview of the selected methods and tools and their relevance to healthcare and toxicology. Next, we describe the  FAIRness assessment for the selected datasets, including both manual method and automated tools. We then present the current state of development of our platform (HazardChat), highlighting key findings and areas for improvement. Finally, we discuss the implications of our findings for data governance and provide insights about our future work.

\section{Methods and tools}
The experience and outcomes reported in this paper have been gained in the context of a research project aiming at understanding how chemicals data from various (governmental and scientific) sources could be used for risk assessment purposes in healthcare. 
To scope this paper and for the purpose of this study, we selected ten freely accessible and widely used chemicals data sources relevant for consumer products and human health risk assessment (see Table \ref{table:db_list}). The intention was not to be exhaustive within the data source selection, but rather to identify recurrent problems from a selection of widely used sources and contribute to solve them. We applied the following set of selection criteria:
\begin{enumerate}
    \item The dataset should be widely adopted by researchers, policy makers or public authorities acting on chemical regulation, risk assessment and mitigation within North America and Europe.
    \item The dataset is publicly accessible to anyone through any internet browser.
    \item The dataset is hosted either in North America or in Europe (for the purpose of scoping of this study. These geographical areas have been active in chemicals data management at the Organisation for Economic Co-operation and Development, OECD, level).
    \item The dataset is relevant for consumer exposure and public health (datasets focused purely on environmental health were not considered).
\end{enumerate}

Note that at the time of writing this paper, the new ECHA CHEM portal is still under development and thus this portal was not analysed. 

We first manually analysed the FAIRness of the datasets. Then we automaticaly evaluated them using the tools FAIR Checker \cite{gaignard2023fair} and F-UJI \cite{huber2021f}.
\href{https://fair-checker.france-bioinformatique.fr}{FAIR Checker} is a free online tool that assesses whether a dataset adheres to the FAIR principles. Users provide a valid persistent identifier (PID) or URL, and the tool scans the dataset's landing page to conduct a comprehensive assessment. Results are visually presented in a radar chart, with normalized evaluation scores ranging from 0 (not satisfied) to 100 (completely satisfied) for each of the four FAIR principles. In addition, a detailed table provides scores, test results, log messages, and recommendations for each test. Notably, FAIR-Checker does not differentiate between data and metadata evaluations.
\href{https://www.f-uji.net}{F-UJI}, another user-friendly tool, automatically evaluates the FAIRness of the datasets. Users enter the dataset's URL, and if online metadata is available, they can specify its type, such as OAI-PMH, OGC CSW, or SPARQL. F-UJI generates a comprehensive report summarizing the assessment results, including a multi-level pie chart that visualizes the dataset's overall FAIRness level. Although the pie chart offers a general overview, it is not interactive. The detailed report delves into each test performed, indicating the corresponding FAIR level (initial, moderate, or advanced) using colored checkmarks (light, medium, and dark, respectively). Debug messages for each test allow users to independently verify and evaluate the test outputs.

\begin{table}[!t]
\caption{List of Selected Databases}
\label{table:db_list}
\centering
\begin{tabular}{||p{3cm}|p{4.5cm}||} 
 \hline
 Publisher & Used URL  \\ [0.5ex] 
 \hline\hline
 ECHA REACH registered substance factsheets & \href{https://echa.europa.eu/information-on-chemicals/registered-substances}{https://echa.europa.eu/information-on-chemicals/registered-substances}  \\ 
 \hline
ECHA Classification and Labelling (C\&L) Inventory & \href{https://echa.europa.eu/information-on-chemicals/cl-inventory-database}{https://echa.europa.eu/information-on-chemicals/cl-inventory-database}  \\
 \hline
 ECHA database for information on Substances of Concern In articles as such or in complex objects (Products) (SCIP) & \href{https://echa.europa.eu/scip-database}{https://echa.europa.eu/scip-database}   \\
 \hline
 European Commission’s Cosmetic ingredient database (Cosing) & \href{https://ec.europa.eu/growth/tools-databases/cosing}{https://ec.europa.eu/growth/tools-databases/cosing/}  \\
 \hline
 Joint Research Centre’s Information platform for chemical monitoring (IPChem) & \href{https://ipchem.jrc.ec.europa.eu/}{https://ipchem.jrc.ec.europa.eu/}  \\
 \hline
 ChemSpider (Royal Society of Chemistry) & \href{https://www.chemspider.com/}{https://www.chemspider.com/}  \\
 \hline
 EPA ChemView & \href{https://www.epa.gov/assessing-and-managing-chemicals-under-tsca/introduction-chemview}{https://www.epa.gov/assessing-and-managing-chemicals-under-tsca/introduction-chemview}  \\
 \hline
 Comparative Toxicogenomics Database (CTD) & \href{https://ctdbase.org/}{https://ctdbase.org/} \\
 \hline
 EPA CompTox Chemicals Dashboard & \href{https://comptox.epa.gov/dashboard/}{https://comptox.epa.gov/dashboard/}  \\
 \hline
 Toxin and Toxin Target Database (Toxic Exposome Database, T3DB) & \href{http://www.t3db.ca/}{http://www.t3db.ca/}  \\  [1ex]
 \hline   
\end{tabular} 
\end{table}

The URL used in this assessment follows the criteria: 1-) The preference is to use the url published in the FAIRsharing; 2-) if not there, use the URL published in PubChem; 3-) if not there, use the URL of the publisher (where the data source can be accessed). \href{https://fairsharing.org/search?fairsharingRegistry=Database}{FAIRsharing.org} \cite{sansone2019fairsharing} is a community-driven resource that promotes the FAIR principles and the use of standards, databases, and policies. It aims to classify and align research data policies across publishers and funders, moderate cross-publisher discussions on repositories, define and register FAIR maturity indicators and metrics, and build guidance and training materials. In their website, we can find the metadata of data sources that follows international standards. Normally, the quality of the metadata is ensured by workgroups of the domain, resulting in better scores from the assessment tools.
PubChem \cite{kim2023pubchem} is another source of metadata for chemical datasets. It is the largest open-access chemistry database. Launched in 2003 and hosted by the NIH, PubChem covers hundreds of millions of compounds, substances, and bioactivities, providing extensive chemical information collected from almost 1,000 data sources. However, the quality of the metadata published in PubChem must be verified by the publisher, what can result in different score when applying the FAIR assessment tools. 
Finally, the publisher website is the place where data can be accessed (browsed or downloaded) by humans. They often are not designed to have computer-interpretable information. Some publisher also provide APIs to access data, as well as unstructured documents (e.g., guides, tutorials, blogs) to describe them. The publisher's website have the lowest score from the assessment tools that we used.

\section{FAIR Analysis}

\begin{figure}%
 \centering
  \parbox{1.2in}{
  \includegraphics[width=1\linewidth, height=3cm]{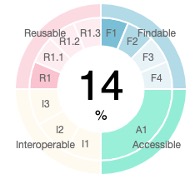}  
  \caption{Score F-UJI}%
  \label{fig:ffuji}}%
  \qquad \begin{minipage}{1.2in}%
  \includegraphics[width=1\linewidth, height=4cm]{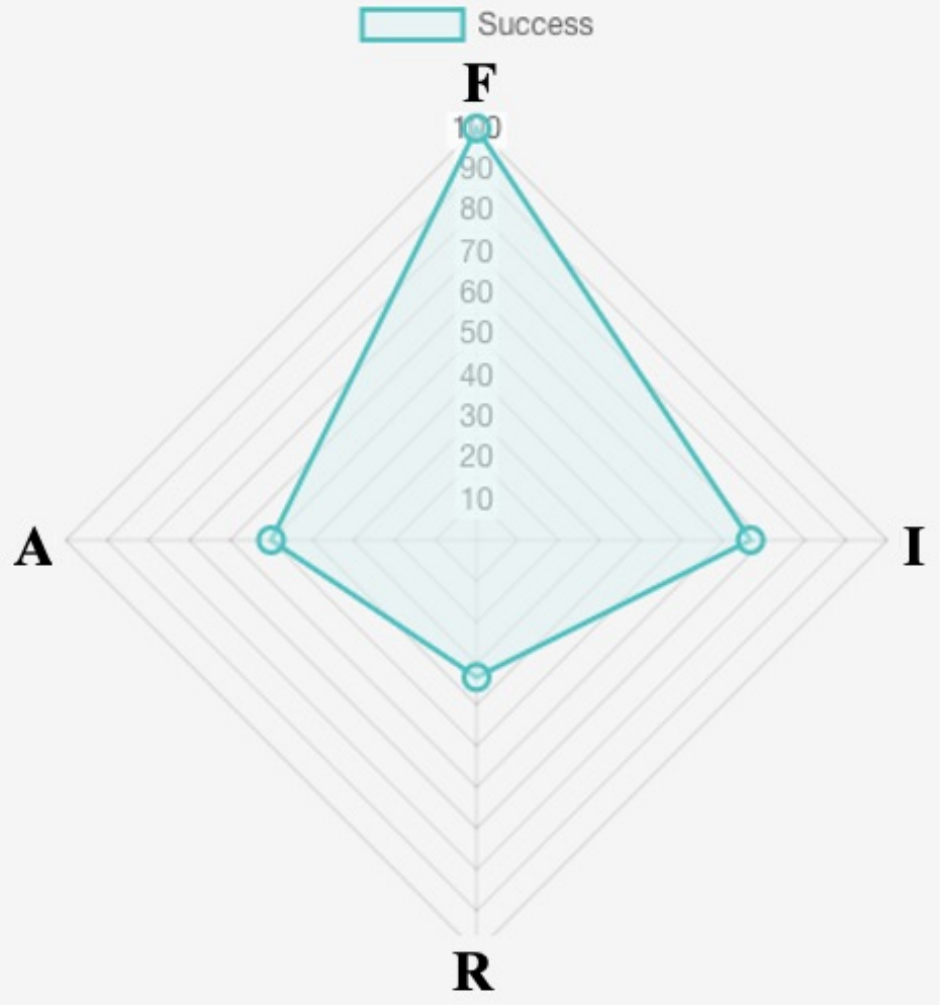}
  \caption{Score FAIR Checker}%
  \label{fig:ffchecker}%
  \end{minipage}%
  \end{figure}

The two FAIR analysis tools selected for this work use different metrics and scoring system, producing different FAIRness evaluation results \cite{sun2022comprehensive}. For instance, F-UJI and FAIR Checker evaluate seventeen and twelve criteria, respectively. This section presents the main observations and the discussions on the results. However, to improve readiness and respect space limit, we do not include all details about the analysis in this paper, but it can be found in \cite{dasilveira2024fairevaluationwidelyused}. Figures \ref{fig:ffuji} and \ref{fig:ffchecker} illustrate the type of graphical outcomes produced by F-UJI and FAIR Checker, respectively, for the REACH database. The manual analysis is not presented in details here neither, but it can also be found in \cite{dasilveira2024fairevaluationwidelyused}. 

The ten selected databases have all a persistent URL, what partially satisfy the findability criterion. Note that the databases managed by ECHA (including Registered Substances, C\&L Inventory, SCIP), EU Commission (CosIng), and ChemView (managed by EPA) were not yet indexed in FAIRSharing nor in PubChem. For this reason, the metadata was not accessible and the accessibility score is lower than ChemSpider, CTD, Comptox, and T3DB.

In terms of accessibility, all datasets satisfied this criterion for manual analysis methods due to their open access nature. Notice that some datasets, such as ChemSpider, CTD, and ChemView, require users to submit an access request to download or reuse the data. But, their websites provide sufficient guidance for users to navigate this process. We found that the accessibility scores were low for automatic analysis as none of the websites provided enough information in the metadata (or do not have metadata) to explicitly indicate where data was published. Additionally, the absence of license information in the metadata further reduced the accessibility scores.

Interoperability emerged as the most challenging criterion to satisfy. Although textual or video documents explaining the dataset structures are sometimes available for humans, machine-interpretable information is often missing. Only three publishers (CompTox, ChemSpider, and ChemView) offer API access and provide documentation for their use. A critical interoperability issue identified was the inconsistent use of CAS numbers\footnote{https://www.cas.org/cas-data/cas-registry} as unique identifiers for chemical substances. For instance, the uniqueness constraint was not always respected, and some substances were published without CAS numbers. Harmonizing vocabularies, identifiers, and descriptors are essential steps to improve interoperability, as well as establishing mappings between different data sources. Initiatives like PubChem, ChemSpider, and IPCHEM have demonstrated the feasibility and benefits of aggregating data from various sources and should receive support from the international community.

The reusability of datasets is generally enhanced by the availability of downloadable data in widely adopted formats such as xlsx, csv, and xml, accompanied by comprehensive documentation and tutorials. While these resources are user-friendly and assist human users in reusing data, they are often inadequate for computer systems. Access to metadata, or at least APIs, is crucial for improving reusability. Despite the valuable efforts of the community to describe and publish metadata on FAIRSharing, often using standards like Dublin Core, much important information is still missing. For instance, license information was frequently absent from the selected metadata. Consequently, the FAIRSharing community proposes minimal descriptions of data, with the expectation that publishers will maintain and complete this data.

In summary, our conclusions are aligned with the Commission Staff Working Document \cite{EUSTAFFdoc2023}, published in December 2023, that highlights the following issues: 
“\textit{1. Chemicals data is scattered 2. Chemicals data is not always interoperable 3. Chemicals data is not always accessible 4. Academic data are insufficiently considered 5. Lack of availability of certain types of chemicals data 6. Not all study results are reported by duty holders 7. Lack of mechanism to identify emerging chemical risks.}“

In our analysis, we concluded that the scattered nature of chemicals data can be addressed by further developing "data aggregator systems" (i.e., IPCHEM, ChemSpider, and ChemView) that offer a unified query/view interface across multiple datasets. To tackle interoperability issues, metadata with a minimal set of standardized information should be made available by all data providers in a common metadata registry (e.g., FAIRSharing.org) with persistent identifiers linking back to the data repositories. Accessibility challenges can be mitigated by ensuring open access to data, without breaching intellectual property and commercial interests. The insufficient consideration of academic data could be improved by implementing a duty to notify (condition to fund projects) on a common chemicals data platform. Additionally, researchers should be trained and incentivized to publish their data according to FAIR principles. Government agencies should also be supported in implementing/checking FAIR data practices. Lastly, to identify emerging chemical risks, we propose to develop methodologies and tools like KGs, and AI-powered tools for data organisation and analysis. 

\section{HazardChat Platform}
In the previous section, we suggested that FAIR data could be easily reused to address problems related to chemical substances, particularly those impacting human health. Here, we demonstrate the advantages of having linked data available to healthcare professionals. To achieve this, we designed and are currently implementing HazardChat, a platform that uses a chatbot interface and exploits aggregated information on chemical hazards collected from several reliable public sources and stored as a knowledge graph. This work focuses on identifying and analyzing the barriers to building a database that disseminates data on hazardous chemical substances commonly used by European industries and found in both the USA and European (EU) markets. We demonstrate the impact that FAIR data can have on building new platforms for specific healthcare needs and discuss the strengths and weaknesses of existing technologies for enhancing data quality and accessibility.

For building the KG, we used three sources of information: 
\begin{enumerate}
\item Chemical factsheet from ECHA REACH database. Collected in November 2023, it was used to extract the list of hazardous substances. Since the information is not available in a downloadable format, it requires parsing the html files. The template is not standard, making the parsing difficult or requiring manual intervention. All substances have an EC number (a European identifier for chemical substances), but not all of them have a CAS number. From the substance factsheet we also extracted the Hazard Class, the hazard phrase used to further describe the class, and the type of products where the substance can be found (Product Category). 
\item EPA CTD database. Collected in June 2024, it was used to extract the disease associated to hazardous substances. The list of substances and the map to the diseases are available in csv and xml formats. The substances are identified by an internal code (ChemicalID) and, when available, the CAS number. The diseases are identified with the MESH\footnote{https://www.nlm.nih.gov/mesh/meshhome.html} code or the OMIM\footnote{https://www.omim.org/} code, when both exist, the preference is given to MESH code. The list of CTD substances is not a subset of the list of REACH registered substances. Thus, we used only the intersection of them to map with diseases.
\item The National Institute for Occupational Safety and Health (NIOSH) database\footnote{https://www.cdc.gov/niosh/npg/npgsyn-a.html}, from the USA Centers for Disease Control and Prevention (CDC). Collected in June 2024, it was used to identify the organs that are impacted by the hazardous substances. The information is presented in the html file describing the substances. Each substance has a CAS number as identifier. After parsing the files 34 distinct organs or systems (e.g., respiratory system, eyes, etc.) were identified. However, CDC also consider as organ parts of the body such as enzymes (e.g. blood cholinesterase). Moreover, beside of the organs' name there is, sometimes, a short explanation/precision between parenthesis. A manual intervention was sometimes necessary to reduce ambiguities and duplication.
\end{enumerate}
The three data sources have complementary information and different properties, but all are open access. The work done allows collecting 21168 hazardous substances, 5825 related diseases, 34 target organs, and 11 hazard classes. By analyzing these three sources of information, a healthcare professional can manually find the answer to the question asked in the introduction section (What are the potential health impacts, particularly on the heart, of exposure to Acrylaldehyde?). The manual search starts by opening the webpage of the substance named Acrylaldehyde, in the factsheet of the REACH dataset, and see if this substance has a Hazard Class (in this work we are only interested on hazardous substances) and take note of the substance CAS number (or any other identifier that can be used later). Optionally, based on the identifier and within the CDC website, the user can check if between the target organs there is `heart'. Finally, the identifier will be used to search in the CTD file for mappings between the substance and the set of diseases. The healthcare professional will select, from this set, only the diseases related to `heart'. This process is time consuming and error prone. Some of these tasks can be automatized and the process can be simplified. One of the proposes of HazardChat is to do it by linking all three data sources via a KG and create a query that correspond to what the user wants. We develop two ways to access the required information: via a graphical interface, or via a chatbot.

\subsection{Graphical interface}

\begin{figure}%
 \centering
  \includegraphics[width=6cm, height=5cm]{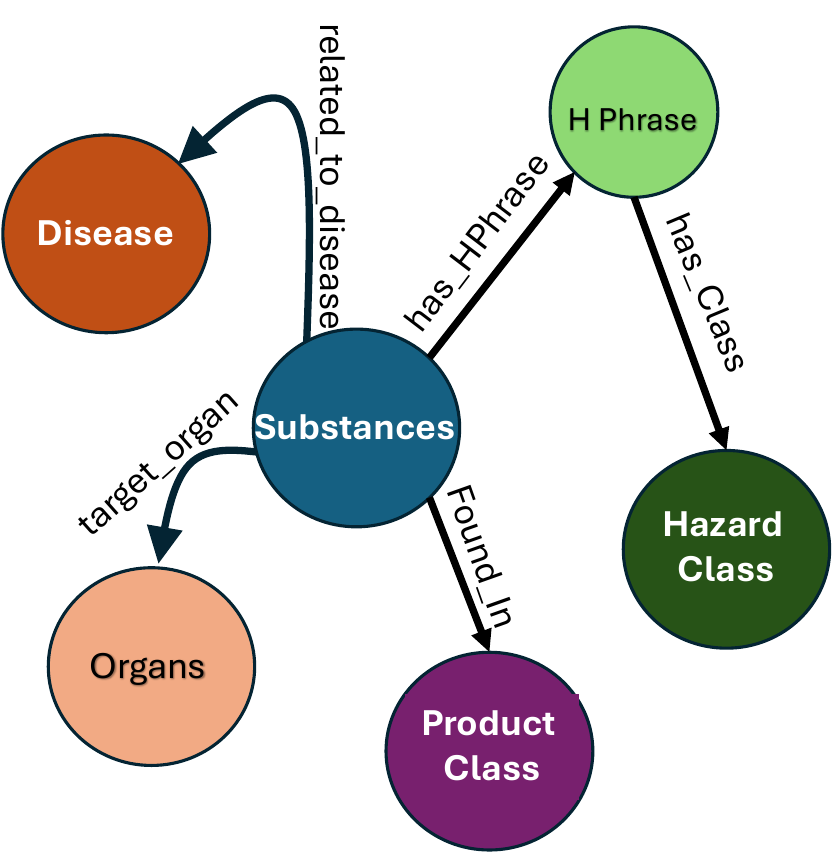}  
  \caption{Schema of the Knowledge Graph}%
  \label{fig:schema}%
\end{figure}

After parsing the data sources and identifying all relevant information, we designed the schema of the database (Figure \ref{fig:schema}). Since this schema can evolve over time, we decided to adopt a graph model that is more flexible and relatively easy to query.
The database adopted to store the data is Neo4J. There are two interfaces to access the information in Neo4J. One is `Explore', that is similar to a SPARQL endpoint and allows writing and executing Cypher queries. The other is Bloom, a data visualization tool to quickly explore and freely interact with Neo4J's graph data platform with no coding required. The team has previous experience with this database and the Bloom interface shows to be appreciated by the end-users (with no ICT background) to explore the KG. Another reason is that Neo4J stores property graphs, what facilitates adding properties to the relations. For instance, to represent a relation between a substance and a disease that exists only if the quantity of exposure to a chemical substance is superior to a certain value. 

\subsection{Chat interface}

The chatbot was built based on the OpenAI model (4o-mini) and implemented in Python. We use our knowledge graph as an external source to apply the retrieval-augmented generation (RAG) \cite{lewis2020retrieval} approach and reduce hallucinations while improving the quality of the answers. The prompt template and the results of the execution of the RAG can be seen in Table \ref{table:prompt}. The prompt creation process follows several steps. First, we defined the required expertise of the model (healthcare professional with strong background on chemistry and toxicology). Second, we provided a set of examples of queries/answers (few-shot approach) selected according to the cosine similarity of the embedded queries. Third, we add in the context of the prompt the schema of the graph. Forth, we asked to explicitly present the explanation of the answer (Chain-of-Thought \cite{wei2022chain}). Fifth, we asked to the model to create the Cypher query that represents the user's question. Sixth, we asked to the model to validate the query before showing it or, otherwise, to say that it does not know the answer. Finally, we also asked to produce a short text explaining/summarizing the outcomes of the query in a natural language, when it is relevant. The user will have access to the query, to the results provided by the graph, and to the text explaining it.  
\begin{table}[!t]
\caption{Example of Adding Information from a KG into the Context of a LLM Prompt.}
\label{table:prompt}
\centering
\begin{tabular}{||p{7.5cm}||} 
 \hline
 \textbf{Prompt template}  \\ 
 \hline
You are an expert on creating Cypher queries. You also have a deep knowledge in Healthcare and Toxicology. Think step-by-step to answer the question. First, given an input question, create a syntactically correct Cypher query to run.      
Here is the graph schema: 
\\… [call db.schema] … \\
If the question matches one of the sample questions in the KG then just use the same query used to answer it.
If the user asks to retrieve a property of an entity of the graph, given its name, then use a WHERE statement and a cypher regular expression matching without case sensitivity, and filter the results by the name of the entity.
Ensure the generated query captures relevant information from the graph database without reducing the retrieved data due to variations or synonyms in user wording.
Use the outcome of the query to answer the user's question. If the question has several answers, list each of them and create a summary to explain the context of the list of the answers. If you do not know the answer, just say I don't know.\\
Below there are some examples of questions and their corresponding Cypher queries and results.
\\… [4 examples are given] … \\
\textbf{User input:} What are the potential health impacts, particularly on the heart, of exposure to Acrylaldehyde ? 
\\ 
 \hline
\textbf{Result } 
\\ 
 \hline
 Acrylaldehyde can potentially impact the heart by causing the following diseases: "heart block", "hypoplastic left heart syndrome", "neurodevelopmental disorder with or without anomalies of the brain, eye, or heart", "arterial occlusive disease, progressive, with hypertension, heart defects, bone fragility, and brachysyndactyly", "heart arrest", "heart valve disease", "heart septal defects, ventricular", "heart-hand syndrome, slovenian type", "heart failure", "heartburn", "heart defects, congenital", "heart injury", "heart failure, diastolic".
\\
 \hline
\textbf{Cypher query generated by the LLM} 
\\ 
 \hline
 $MATCH (o:Organ \{Organ: 'heart'\}) \leftarrow$ [:target\_organ]- $(sub:Substance \{name: 'Acrylaldehyde'\}) -[:related\_to\_disease] \rightarrow 
 (d:Disease)$ where toLower(d.DiseaseName) contains $'heart')$
RETURN d.DiseaseName
\\
 \hline \hline
\end{tabular} 
\end{table}

\begin{figure}%
 \centering
  \includegraphics[width=8cm, height=5cm]{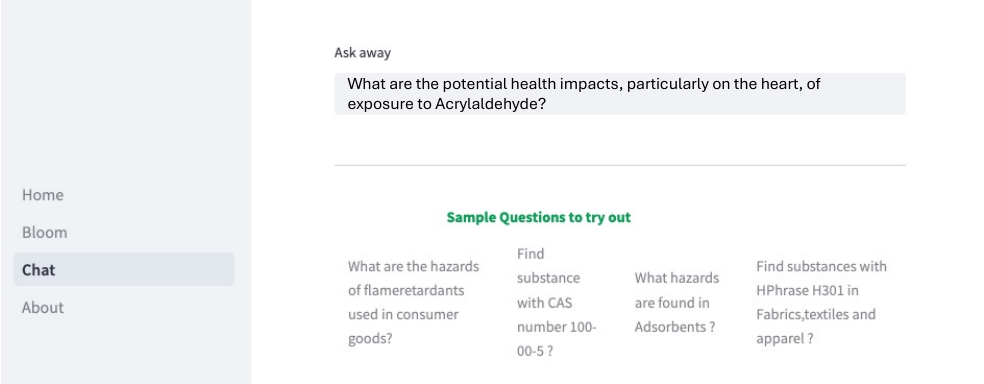}  
  \caption{Interface for the Chatbot}%
  \label{fig:chat}%
\end{figure}
However, more work is needed to understand how performance varies across different ranges of question types, data types, and dataset sizes, as well as to validate our approach with end users. As it is now, HazardChat is a demonstration tool that intend to provide information to healthcare professional. This information must be taken with caution when analysing the patient conditions. 

The process of creating this knowledge graph revels several opportunities to improve the datasets. The format of the published information (html, csv, xml) required a case-by-case solution and human intervention to understand the templates or fields within the files and extract the relevant information. There is a need for a common terminology to name tags/fields (e.g., Substance, Chemical, ChemicalName are used to refer to the same information) in the data sources. Our recommendation is the creation of metadata where the fields are mapped to ontologies of the domain. Another challenge was the identification of unique substances. CAS numbers is the most widely identifier used by practitioners, but we notice that they are incomplete and ambiguous. Moreover, they are not freely open, a paid license is needed to register substances. Several other international initiatives propose a global substances ID, but there is no consensus on what use and the adoption of an unique global ID is far from being the reality. Our recommendation is to have regulators involved in the decision process in order to define common rules and enforce industries and academics to provide a minimum set of information that allows identifying unambiguously a substance or a mixture/product. Finally, the access to the information is rarely provided via APIs with a set common interface. Our recommendation is to have APIs available for each data source. A complementary work that analyzes the EU regulations on this topic and provide recommendations to improve it can be found in \cite{freudenthal2024unlocking}.
With the implementation of our platform, we intend to show the positive impact of having FAIR data, and we list the challenges that still need to be addressed followed by the lessons learned and/or recommendations to solve them.

\section{Discussions}

The combination of FAIR data principles, Knowledge Graphs, and Large Language Models presents a transformative paradigm for the field of knowledge management, specially in critical situations where decisions must be taken quickly. By combining these technologies and principles, healthcare professionals can significantly enhance data accessibility without requiring deep technical skills, thereby, or executing time consuming data search tasks, accelerating medical discovery and improving diseases diagnosis and treatments.

The FAIR principles provide a foundational framework for ensuring that data is findable, accessible, interoperable, and reusable. Adherence to these principles are essential for the construction and enrichment of KGs. Watford et al. \cite{watford2019progress} emphasized the necessity of data interoperability, for instance, in advanced computational toxicology, highlighting the need for robust data management practices to support the development of FAIR-compliant data repositories. The authors also highlight that while current data publishers provide functional and interactive platforms for accessing toxicological information, they often lack interoperability (e.g., using proprietary format, local IDs, no metadata). Improving this requires rigorous data management and stewardship practices, thereby enhancing the ability to integrate and utilize diverse data sources effectively. The EU Chemicals Strategy for Sustainability reinforce this idea in the communication published in \cite{CSS2020} where they request “\textit{[…] free the data access of technical or administrative obstacles, according to the principles that data should be easily findable, interoperable, secure, shared and reused by default. Data will be made available in appropriate formats and tools […] - to ensure interoperability.}”. 

With this opportunity in mind, funding agencies like the European Commission have started to incentivise the scientific community to improve the FAIRness of research and data (including descriptions of methods used) and publish them as open access. Recently initiatives have taken place in FAIRness adoption (\href{https://www.go-fair.org/}{Go FAIR}), also within the chemicals and materials sector (\href{https://www.nanosafetycluster.eu/nsc-overview/nsc-structure/working-groups/wgf/}{NanoSafetyCluster}). Furthermore, a first generation of tools to manually assess FAIRness (e.g., \href{https://satifyd.dans.knaw.nl/}{DANS}, \href{https://satifyd.dans.knaw.nl/}{SATIFYD}, \href{https://ardc.edu.au/resource/fair-data-self-assessment-tool}{ARDC}) were developed and made accessible in the form of online questionnaires. The second generation implemented an automatic assess process (e.g., \href{https://www.f-uji.net}{F-UJI} and \href{https://fair-checker.france-bioinformatique.fr}{FAIR Checker}). Our work demonstrate that there are advances in this practice, but there is also room for improvements.
Complementary to FAIR data sharing, Knowledge Graphs have emerged as a powerful tool for managing and integrating complex data sets, particularly in the domain of hazardous chemical management. By capturing semantic relationships between entities, KGs facilitate efficient data exploration, querying, and reasoning. Zheng et al. \cite{zheng2021knowledge} showcased a method for constructing KGs by collecting data from unstructured documents through a deep learning-based entity recognition system. This approach aims to create a KG that contains information about the effective risks linked to hazardous chemicals. The KG is made available, but they did not follow the FAIR principles, neither used chatbot to facilitate access to the information. KGs have also been used for real-time analytics and support in diagnosing chemical exposure  \cite{shin2022knowledge}. Over a thousand major chemical substances are represented in their KG. They enable rapid querying and reasoning about candidate substances, facilitating quick diagnosis and early response to chemical exposures at accident sites. But they did not use LLMs, neither followed FAIR priniciples to publish the KG. A notable application of KG + LLMS is the TRSRD database \cite{wang2023trsrd}, which is used to search for risky substances in tea. It illustrates the versatility and utility of KGs to combine diverse sources into one (centralized or distributed) graph. The visualization and the graphical exploration of KGs makes this technology beneficial for quickly discovering relations between entities that are not evident to see in the tabular format. However, deeper analysis of the graph requires some technical background to write queries in specific languages (e.g., Cypher or SPARQL). 

Large Language Models came to remove some technical barriers and facilitate querying data sources. Characterized by their ability to process and generate human-like text, these models can be used to leverage the extraction and summary of information, as well as to generate hypotheses and insights from texts and graphs. The adoption of LLM in chemistry and healthcare domains is rising. Approaches such as ChemCrow and Coscientist \cite{m_bran_augmenting_2024, boiko_autonomous_2023}
combine LLMs with domain tools to augment it capacity of observing and/or acting in the real world for accomplishing tasks that are out of reach of a LLM alone. But, they do not store, share or reuse the produced data. 
For that, Pan et al. \cite{pan2024unifying} proposes to use knowledge graphs to help LLMs further enhance the accuracy of their responses through retrieval-augmented generation. Chen et al. \cite{chen2023enhancing} demonstrated the use of LLMs in conjunction with KGs to enhance emergency decision-making by providing evidence-based recommendations. But the FAIR aspect is not mentioned. A comprehensive review of LLMs and autonomous agents in chemistry \cite{ramos2024review} highlights their ability to act as interactive encyclopedias, retrieving and reasoning about chemical properties without requiring additional fine-tuning. A variety of libraries to explore graph databases with LLM are supported by platforms such as \href{(https://github.com/langchain-ai/langchain)}{LangChain} and \href{https://github.com/run-llama/llama_index}{LlamaIndex}, promoting the emerging of more general class of graph-based RAG applications \cite{edge2024local}, including systems that can create and reason over knowledge graphs.

The intersection of FAIR principles, Knowledge Graphs, and Large Language Models represents a promising frontier for improving data management and utilization in chemical safety and toxicology. We highlight in this work the challenges that still need to be addressed, but by ensuring that data is FAIR, researchers and practitioners can more easily access and integrate diverse data sets, having a more complete overview of the domain. Knowledge Graphs was selected by its capacity to organize data in a flexible and interconnected format, enabling more effective risk management and the prediction of response to chemical exposures. The integration of LLMs further enhances these capabilities by allowing for natural language interaction and data retrieval-augmentation over graph-based data. Together, these technologies hold the potential to transform how we manage and utilize scientific data, leading to improved outcomes in chemical safety, toxicology, and beyond. As these fields continue to evolve, ongoing research and development will be crucial in addressing remaining challenges and fully realizing the potential of these integrated approaches.

\section{Conclusions}
In this paper, we analyzed ten data sources on hazardous chemical substances to assess their compliance with FAIR principles, identify the challenges in aggregating them, and explore the benefits of providing healthcare professionals with a user-friendly interface for querying and exploring this data. The ultimate outcome of this work is a platform built on top of a knowledge graph that offers visual and natural language interfaces for quick access to relevant data, aiding in decision-making regarding the diagnosis or treatment of patients' health issues. We conclude that the current state of these data sources does not fully satisfy all FAIR principles and human intervention to correctly aggregate information from multiple sources is still mandatory. Additionally, crucial information, such as licensing details, are not provided in a structured format. Despite the efforts of some data publishers, there is still a need for consensus on defining a global and unique identifier for chemical elements and mixtures. As a result, the mapping of chemical elements from the analyzed sources cannot be automated due to ambiguities in their identities. Finally, we demonstrate the feasibility of a platform that integrates knowledge graphs, chatbots, and FAIR data to provide easy and quick access to important health-related information. In our future work, we will focus on improving the interpretation of natural language queries, enhancing the precision of our Retrieval-Augmented Generation (RAG) approach, and strengthening the connection with scientific literature.

\section{Acknolwedement}
The authors are beneficiary of an \textbf{AXA Research Fund} postdoctoral grant

\bibliographystyle{IEEEtran}
\bibliography{FAIR_KG_LLM.bib}

\end{document}